\documentclass[12pt,preprint]{aastex}
\usepackage{gensymb} 
\usepackage[version=4]{mhchem}
\usepackage{amsmath}
\usepackage[colorlinks=true,linkcolor=blue,citecolor=blue]{hyperref}
\usepackage{contcaption,mathtools}
\shorttitle{Winged radio sources}
\shortauthors{Pal and Kumari}

\begin{document}

\title{Winged Radio Sources from LOFAR Two-metre Sky Survey First Data Release (LoTSS DR1)}

\author{Sabyasachi Pal}

\affil{Indian Centre for Space Physics, 43 Chalantika, Garia Station Road, 700084, India\\
Midnapore City College, Kuturia, Bhadutala, West Bengal, 721129, India}

\and

\author{Shobha Kumari}

\affil{Midnapore City College, Kuturia, Bhadutala, West Bengal, 721129, India}

\begin{abstract}
Winged radio sources are a small sub-class of extragalactic radio sources which display a pair of low surface brightness radio lobes, known as `wings' aligned at a certain angle with the primary jets. Depending on the location of wings, these galaxies look like {\it X} or {\it Z} and are known as {\it X}-shaped Radio Galaxy (XRG) or {\it Z}-shaped Radio Galaxy (ZRG). We report the identification of 33 winged radio sources from the LOFAR Two-metre Sky Survey First Data Release (LoTSS DR1) out of which 21 sources are identified as {\it X}-shaped radio galaxies and 12 as {\it Z}-shaped radio galaxies. Optical counterparts are identified for 14 XRGs (67 per cent) and 12 ZRGs (100 per cent). We studied various physical parameters of these sources like spectral index, radio luminosity and power. The radio spectrum of the majority of XRGs and ZRGs is steep ($\alpha_{1400}^{144}>0.5$), which is typical of lobe dominated radio galaxies. The statistical studies are done on the relative size of the major and minor axes and the angle between the major axis and minor axis for XRGs.
	
\end{abstract}


\section{Introduction}
\label{sec:intro}
In typical radio galaxies with optical counterparts, there exist two oppositely directed jets (primary jets) coming out from both sides of the optical host galaxy. Winged radio galaxies are a very unique and interesting sub-class of radio galaxies in which a pair of secondary jets align with a certain angle with the primary jets. These secondary jets originate in two different ways. In some cases, it originated from near the central core of the galaxy and aligns with the primary jets, gives rise to an {\it X}-shaped structure and named as {\it X}-shaped radio galaxies (XRGs). In other cases, it originates close to the edges of the primary jets and gives rise to {\it Z}-shaped structure and named as Z-shaped radio galaxies (ZRGs).

3C 272.1 is the first identified winged source that has {\it Z}-shaped morphology \citep{Ri72}. \citet{Ek78} classified NGC 326 as an {\it X}-shaped radio source. For the first time, 
\citet{Le92} catalogued 11 such sources. With the help of the Very Large Array (VLA), Faint Images of Radio Sources at Twenty cm (FIRST) survey at 1.4 GHz, \citet{Ch07} presented a sample of 100 winged sources. \cite{Pr11} presented a list of 156 XRG candidates by automated morphology detection process. Recently a list of 290 XRG candidates is presented in \cite{Ya19}. \citet{Be20} presented a large number of winged sources using the VLA FIRST survey at 1.4 GHz. They detected 296 winged sources of which 161 are identified as XRGs and 135 as ZRGs. Using The TIFR GMRT Sky Survey (TGSS) at 150 MHz, \citet{Bh20} detected 92 winged sources out of which 68 are XRGs and 24 are ZRGs.

The origin of the formation of wings (secondary jets) is very mysterious and is required a very deep study. Although there exist some possible mechanism that explains some of these structure of radio galaxies there are no unique models which fit with all of the winged radio galaxies.

We pointed out some widely discussed models---\\ (i) Backflow of plasma \citep{Le84, Ca02}: This model suggests that the formation of wings in winged radio galaxies are due to the backflow of plasma from the central hotspot and this backflow occurs when jet material is supposed to be released by the hotspots and then streaming back towards the host galaxy.
(ii) Blackhole merger \citep{Me02, Go03, Zi05}: Two supermassive black holes merger is another possibility for the formation of wings \citep{Me02}.
When a relatively smaller black hole is deposited through a galaxy merger process the spin axis of the larger black hole goes under a sudden change.
This change in the spin axis may be the reason for producing the secondary lobes for the radio galaxy \citep{Go03, Zi05}. (iii) Precession of jets \citep{Ro01, De02}: According to this model, it is believed that the secondary lobes are the remnants that are leftover from a rapid realignment of the central supermassive black hole-accretion disk system \citep{De02}.
 it can also be said that the wings are due to the relic emission of the previously active lobes \citep{Ro01} whose direction of emission has been changed with time.
(iv) Twin AGN \citep{La05, La07}: According to this model the formation of wings is due to two pairs of jets are believed to be ejected from two unresolved AGNs in a different direction.

In section \ref{sec:method}, we discuss the data selection and the method of identifying the winged sources.
In this section, we also define how we separate the XRGs and ZRGs from the total winged candidate list.
Section \ref{sec:result} contains the result of our search that includes the properties of XRGs and ZRGs.
The discussion and conclusion are presented in section \ref{sec:disc} and section \ref{sec:conc} respectively.

Through out the work, we have used the following cosmology parameters: $H_0 = 67.4$ km s$^{-1}$ Mpc$^{-1}$, $\Omega_m = 0.315$ and $\Omega_{vac} = 0.685$ \citep{Ag20}.
 
\begin{figure*}
\vbox{
\centerline{
\includegraphics[height=4.2cm,width=4.2cm]{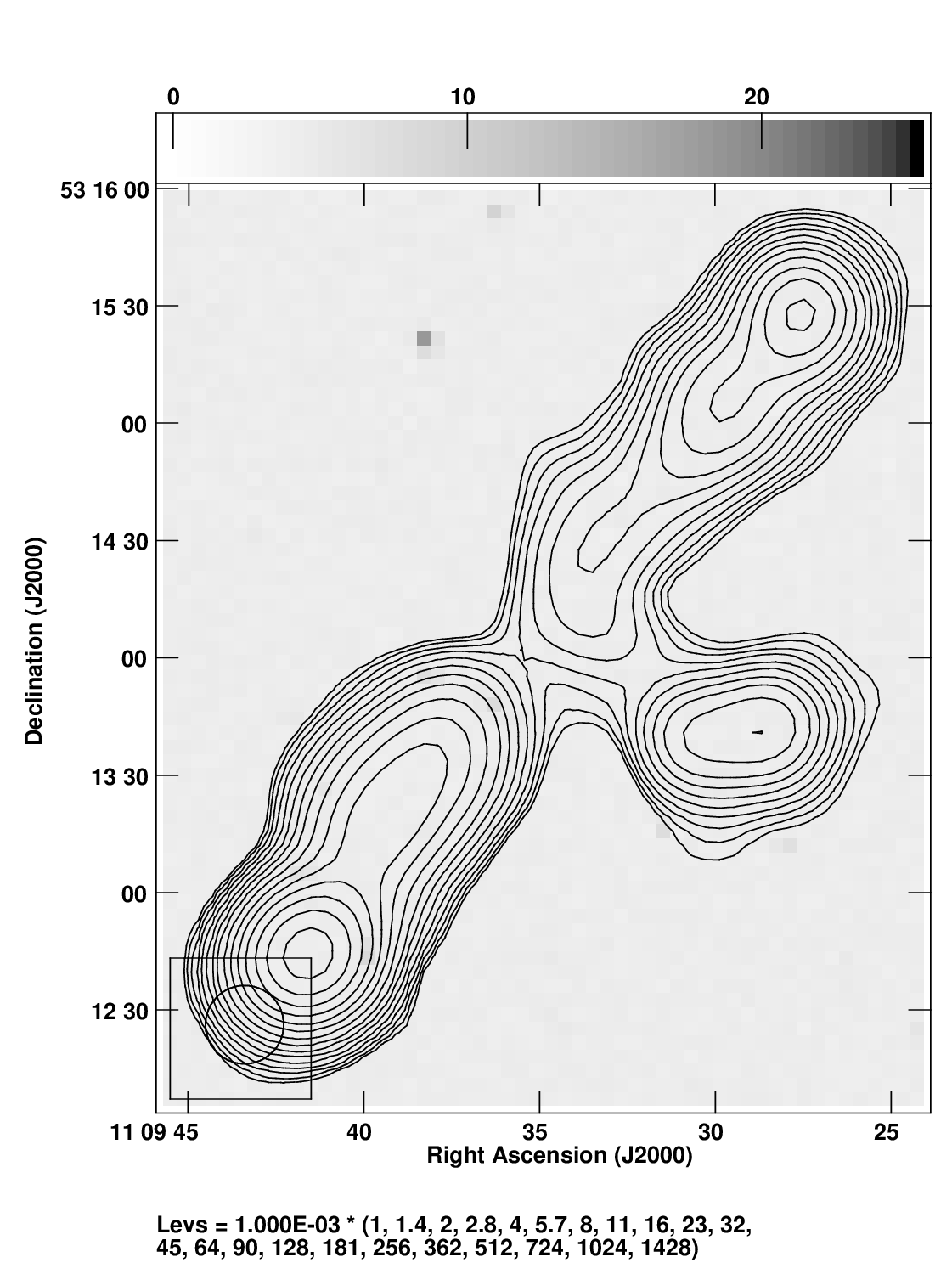}	
\includegraphics[height=4.2cm,width=4.2cm]{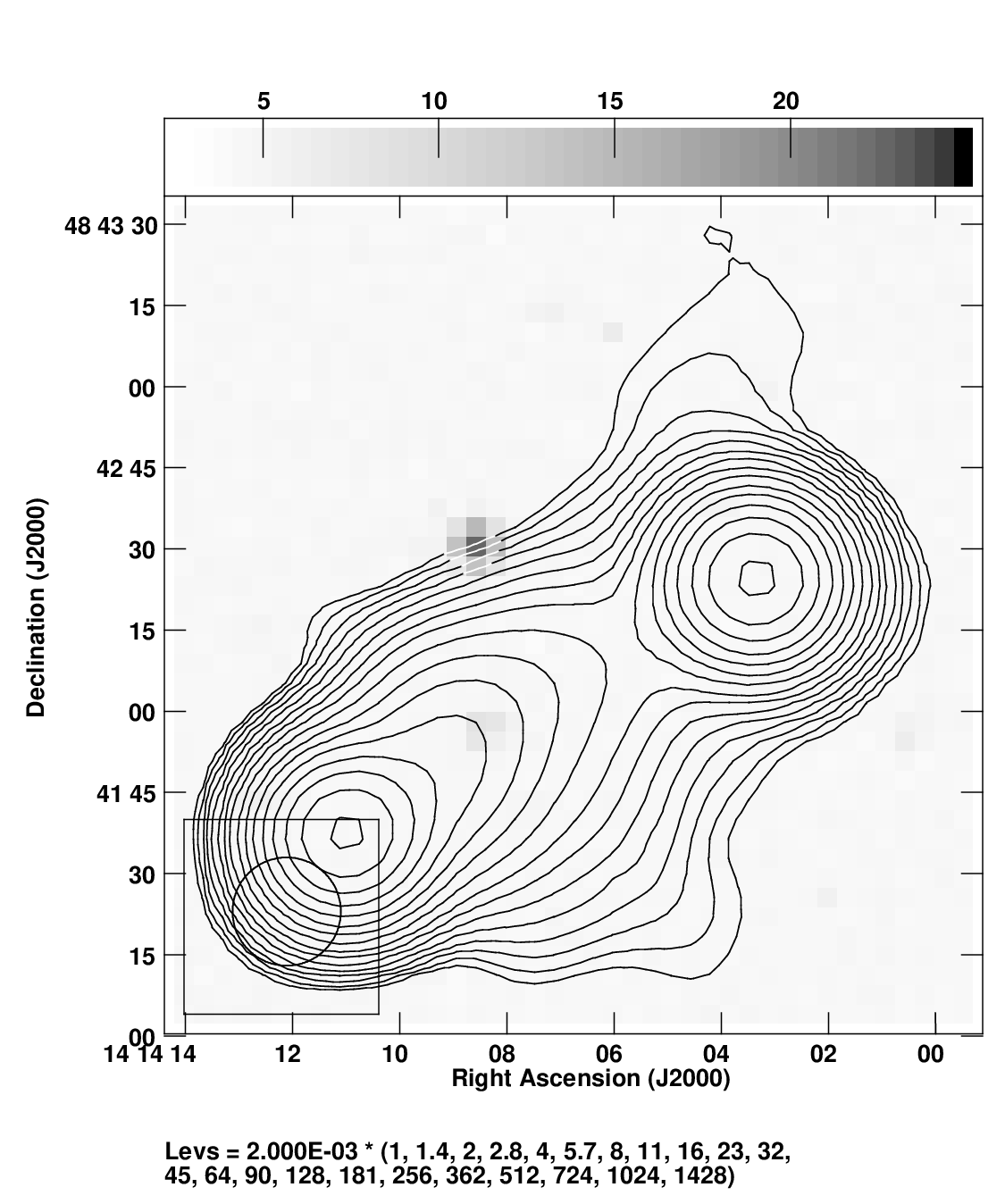}
\includegraphics[height=4.2cm,width=4.2cm]{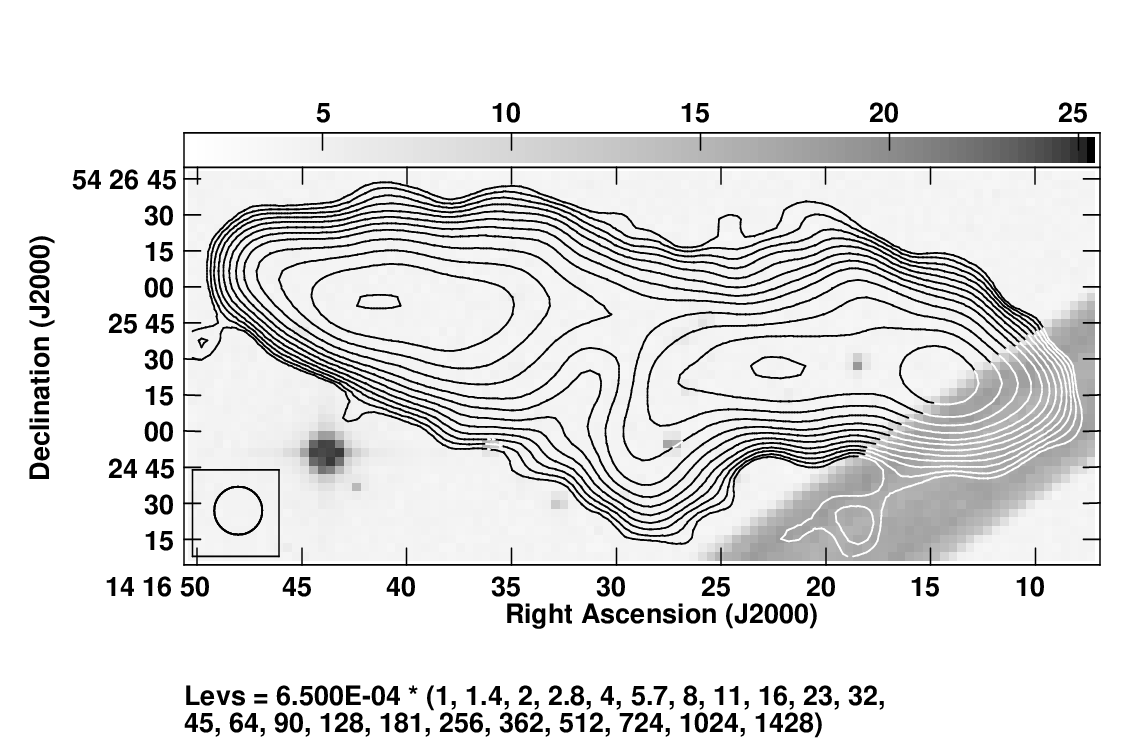}
}
}
\vbox{
\centerline{
\includegraphics[height=4.2cm,width=4.2cm]{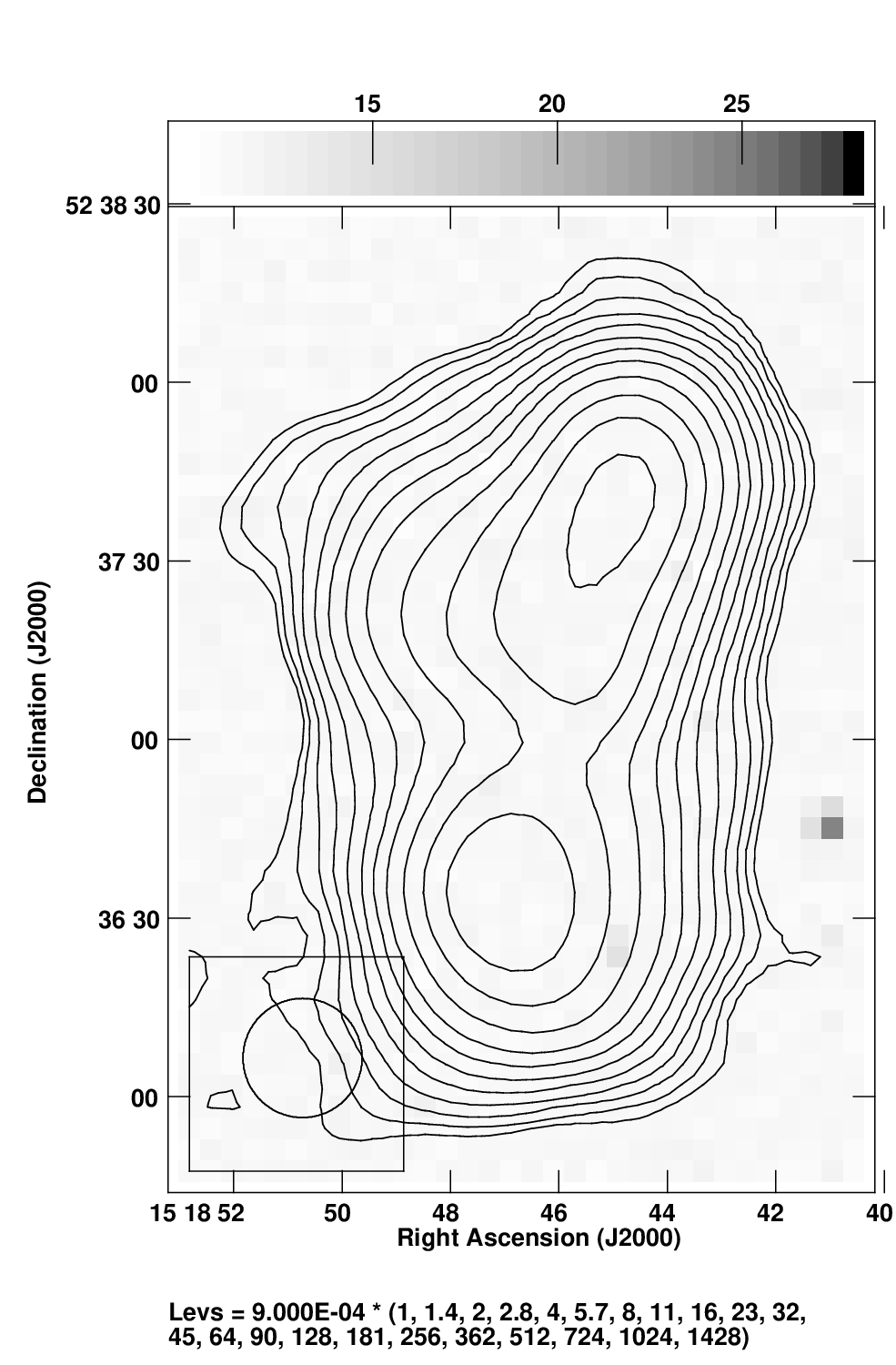}
\includegraphics[height=4.2cm,width=4.2cm]{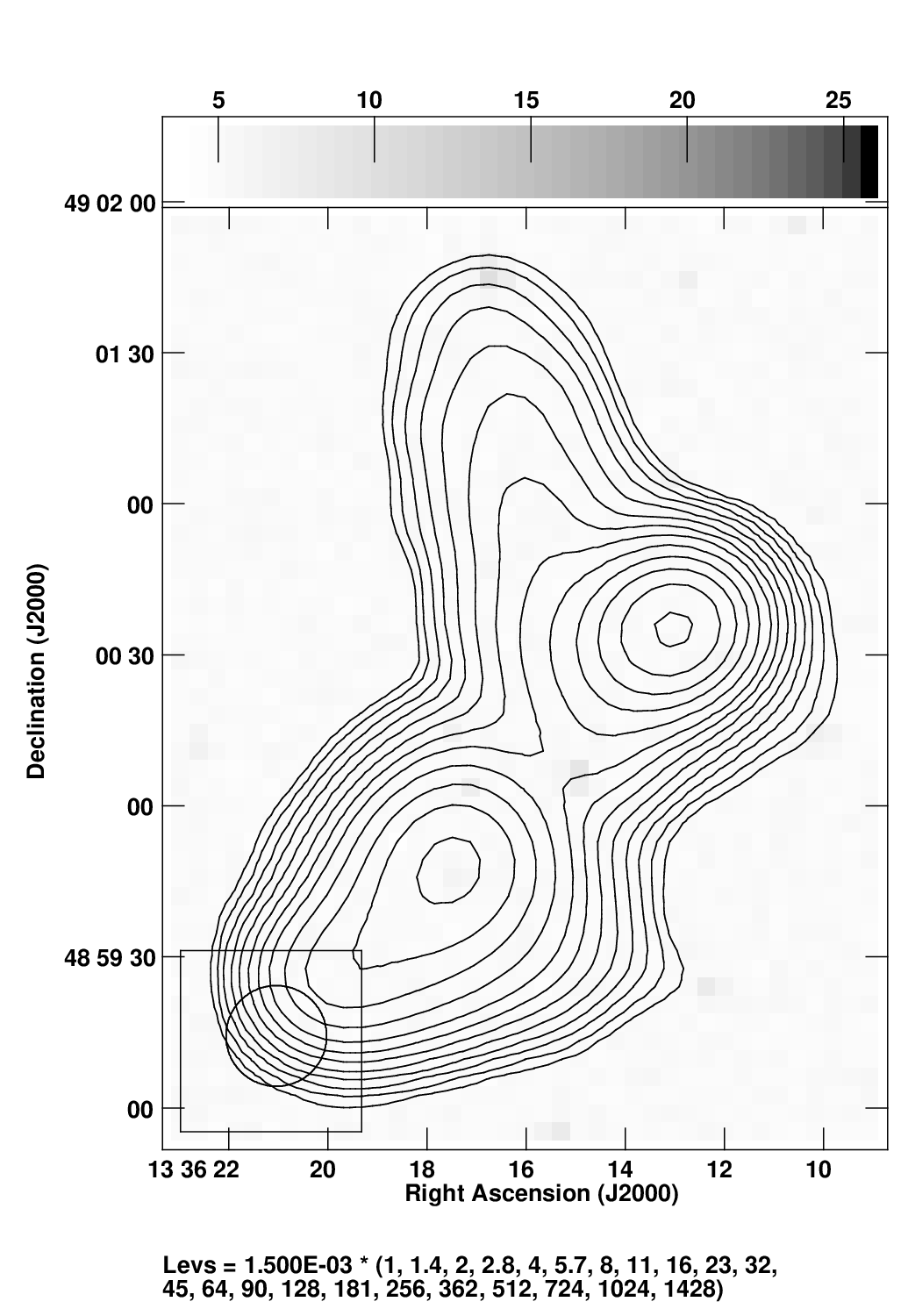}
\includegraphics[height=4.2cm,width=4.2cm]{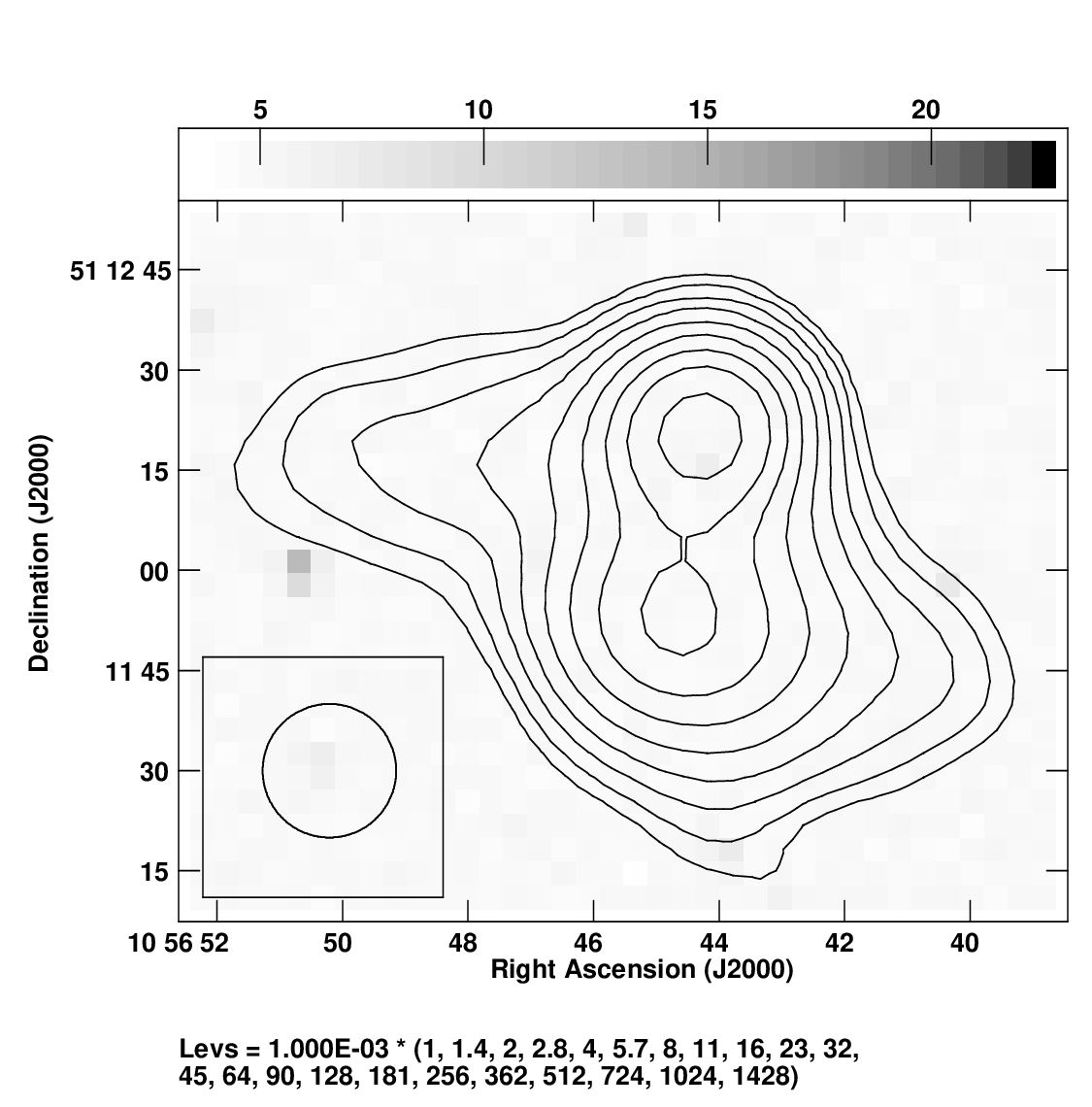}
}
}
\caption{A sample of six LOFAR images of the {\it X}-shaped radio source (contours) overlaid on the DSS2 red image (grey scale).
The relative radio contours are drawn from a selected lowest level (depending upon the local noise) with a increment of $\sqrt2$.}
\label{fig:sample_xrg}
\end{figure*}

\section{Data Selection and Methodology}
\label{sec:method}
\subsection{Definition of XRGs and ZRGs}
The secondary lobes or the less brighten lobes of radio galaxies are known as wings. Depending on the position of the wings, the winged radio galaxies could be classified into two distinct groups -- {\it X}-shaped radio galaxies (XRGs) and {\it Z}-shaped radio galaxies (ZRGs). For XRGs, secondary lobes appear to come from the central spot or near the central region where the near central region is defined as the region covered by $\sim$25 per cent of the primary jet length from the central core. If the wings appear to come from the non-near central area or the edges of the primary jet, the sources are classified as ZRGs. The morphology of sources depends on the radio contour levels (especially the lowest contour) and the signal to noise. We started contours from 900 $\mu$Jy for all images to make a uniform study.

When the secondary lobes appear to come out from the central core or near the central region, we categorize those sources as {\it X}-shaped sources.
Here we define the near central region as the $\sim$25 per cent from the morphological center \citep{Be20}.
On the other hand, when the wings are spotted at the edges of the primary lobes, those sources are catalogued as {\it Z}-shaped sources.

\subsection{The Obsevational Data: LoTSS DR1}
For the present paper, we use LOFAR Two-metre Sky Survey first data release (LoTSS DR1) \citep{Sh19}. LoTSS \citep{Sh17} is a sensitive, high-resolution 120 --168 MHz imaging survey that will eventually cover the entire northern sky and currently, observation is $\sim$20 per cent complete. 
The LOTSS DR1 covers the sky in the region of the Hobby-Eberly Telescope Dark Energy Experiment (HETDEX) Spring field region.
It covers an area of 424 square degrees with a span of right ascension from 10h45m00s to 15h30m00s and declination from 45$^{\circ}$00$^{\prime}$00$^{\prime\prime}$ to 57$^{\circ}$00$^{\prime}$00$^{\prime\prime}$.
With a median sensitivity (S$_{144}$)=71 $\mu$Jy beam$^{-1}$ the survey has a resolution of $\sim$6$^{\prime\prime}$\citep{Sh19}.
Due to the long integration time on each survey grid pointing and the extensive range of baseline lengths in the array, the LoTSS can probe with a combination of resolution, sensitivity, depth, and area to a wide range of angular scales.
LoTSS is 50--1000 more sensitive and 5--30 times higher in resolution in comparison to other low-frequency wide-area surveys, like the TIFR GMRT Sky Survey alternative data release (TGSS; \citep{In17}), LOFAR Multifrequency Snapshot Sky Survey (MSSS; \citep{He15}), GaLactic and Extragalactic All-sky MWA (GLEAM; \citep{Wa15}) and the Very Large Array Low-frequency Sky Survey Redux (VLSSr; \citep{La14}). Taking advantage of better resolution and sensitivity in LoTSS DR1, a large number of head-tail radio galaxies are discovered using the survey \citep{Mi19, Pa21}.

\subsection{Search Method}
There is a total of 325,694 radio sources in LOTSS DR1 \citep{Sh19} with a 5$\sigma$ detection limit. To search for winged radio galaxies, we filtered only those radio sources which have a size more than 12$^{\prime\prime}$, i.e. at least twice the convolution beam size.
Filtering of sources with a size more than 12$^{\prime\prime}$ results in 18,500 sources.
We looked at the fields of all 18500 sources visually to see if there were any new candidate winged radio galaxies. We excluded all the sources from our list which was previously identified from earlier surveys \citep{Ch07, Ya19, Be20, Bh20}.

\subsection{Optical counterpart identification}
We searched for the optical counterparts for each of the newly discovered XRGs and ZRGs using the Sloan Digital Sky Survey (SDSS) data catalogue \citep{Gu06, Al15}. The identification of the optical/IR counterpart of the winged radio galaxy was based on the position of the optical/IR source relative to the radio galaxy morphology. SDSS images are overlaid with LOTSS DR1 images. Since optical host galaxies are much compact than the corresponding radio counterpart, the position of the optical/IR counterpart of XRGs and ZRGs are used as the positions of these sources, when available. The 3rd and 4th columns of Table \ref{tab:X-shaped} and Table \ref{tab:Z-shaped} show the position of these galaxies. When no clear optical/IR counterparts are available, we used the location of the core of the radio galaxy or the intersection of both radio lobes as the position of the galaxy. Out of a total of 21 sources, optical/IR counterparts are found for 14 sources for XRGs (66 per cent). For ZRGs, optical/IR counterparts are found for all of the 12 sources.

\begin{figure*}
\vbox{
\centerline{
\includegraphics[height=4.2cm,width=4.2cm]{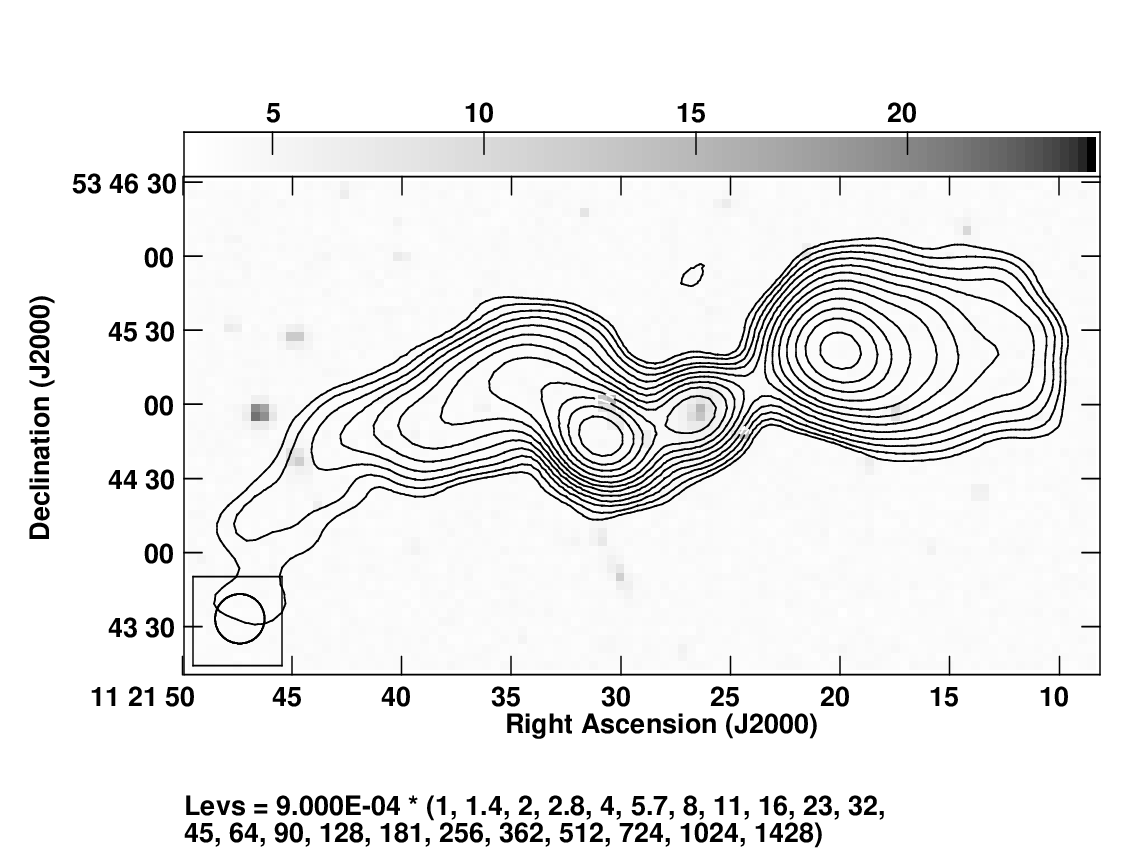}
\includegraphics[height=4.2cm,width=4.2cm]{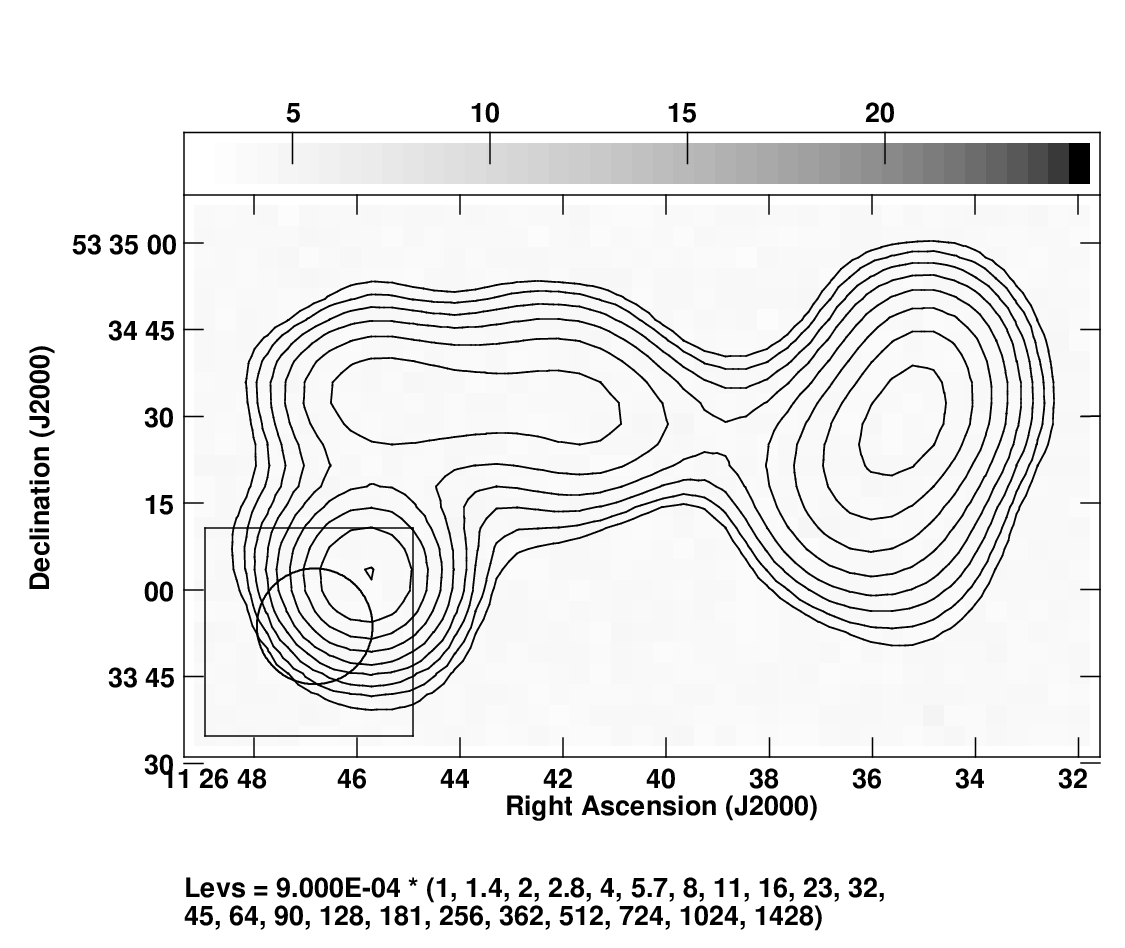}
\includegraphics[height=4.2cm,width=4.2cm]{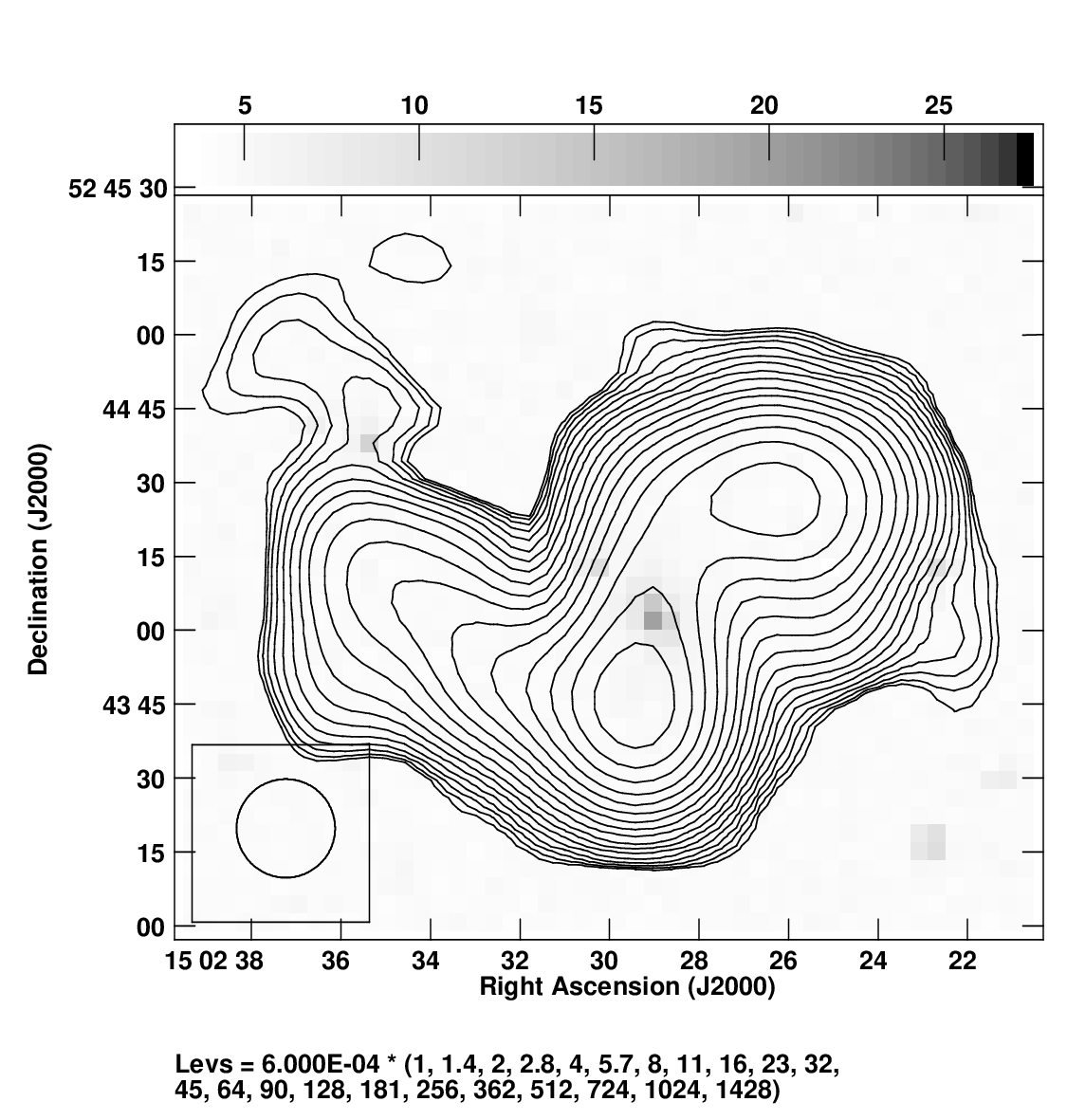}
}
}
\vbox{
\centerline{
\includegraphics[height=4.2cm,width=4.2cm]{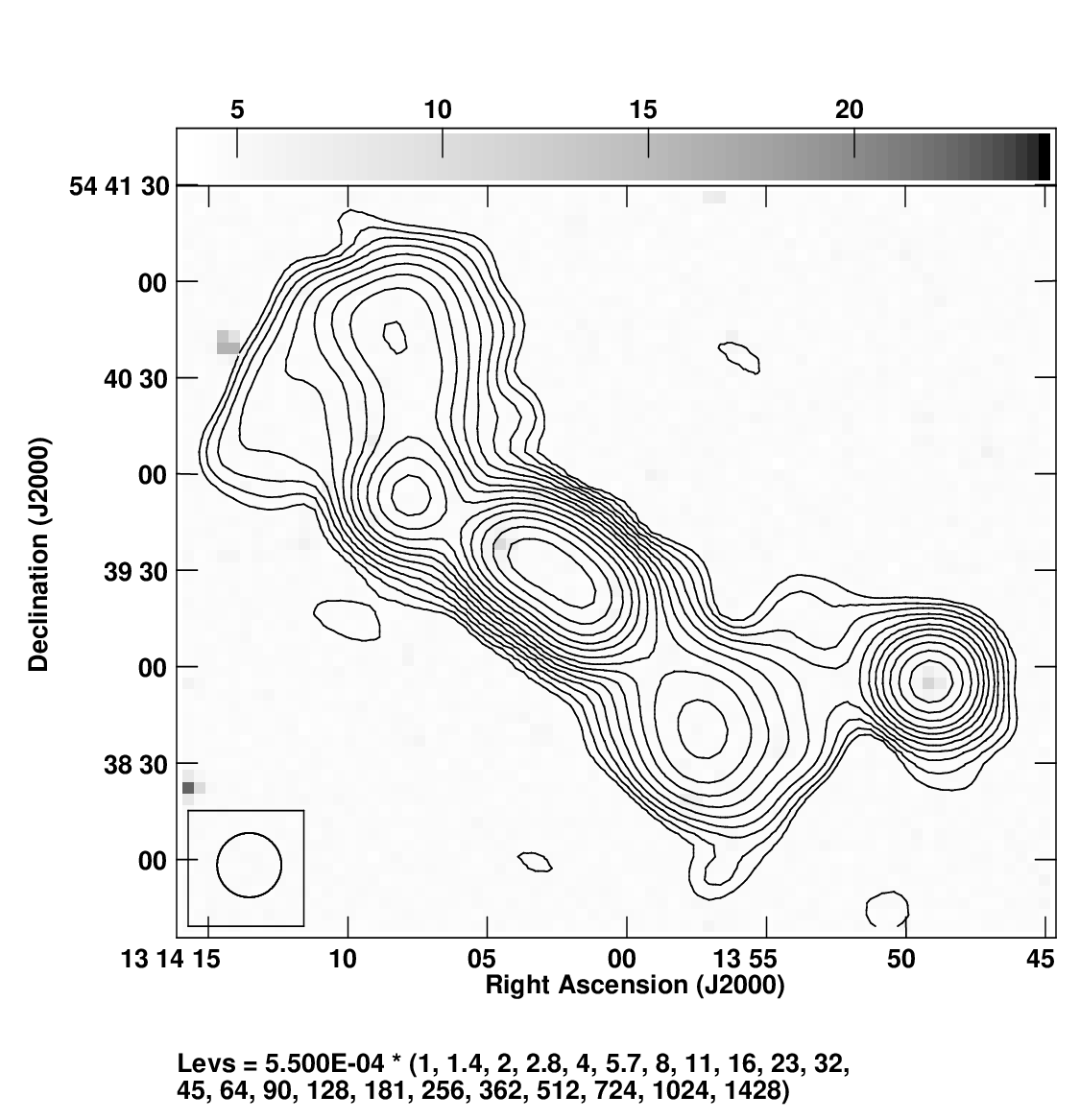}
\includegraphics[height=4.2cm,width=4.2cm]{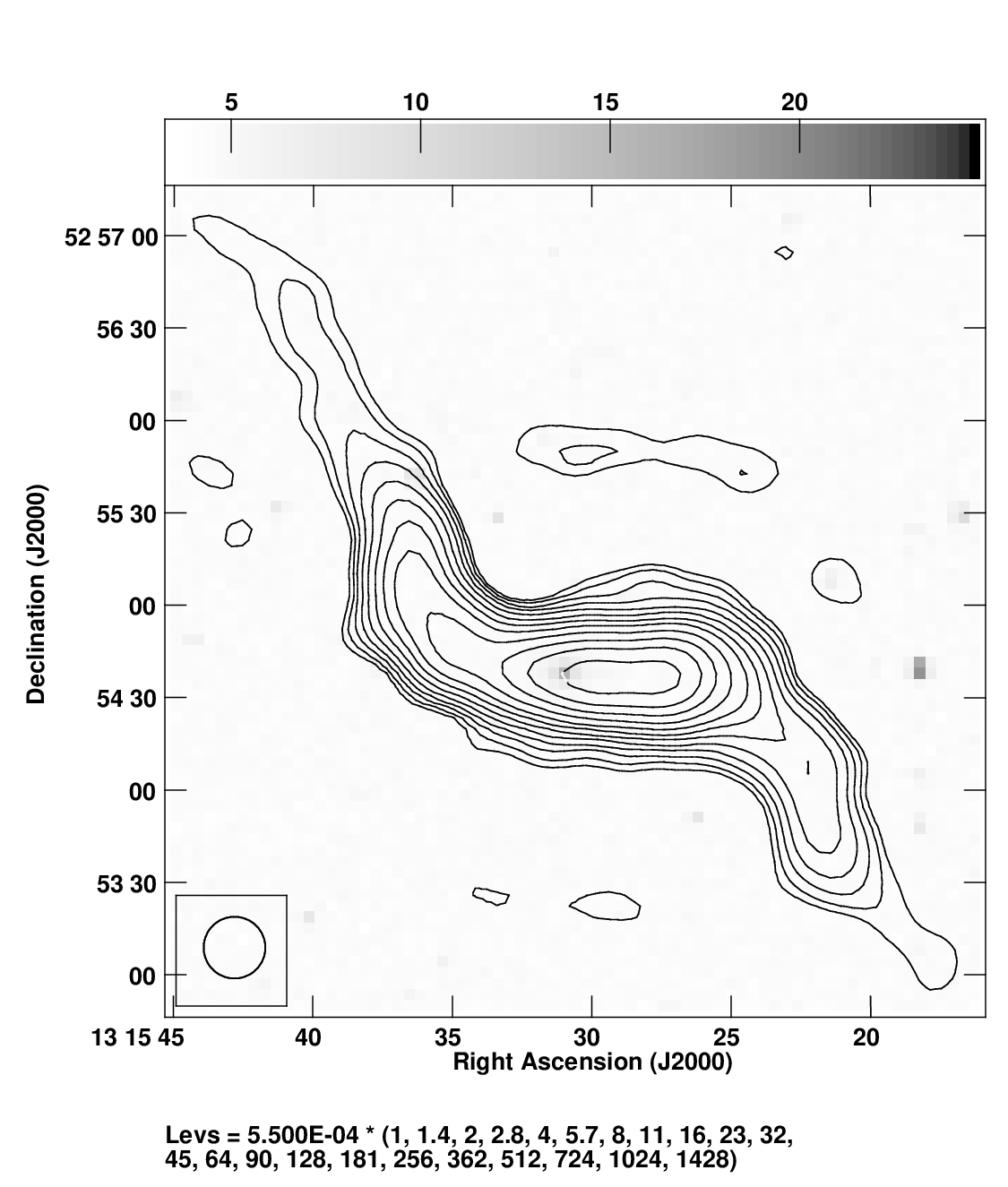}
\includegraphics[height=4.2cm,width=4.2cm]{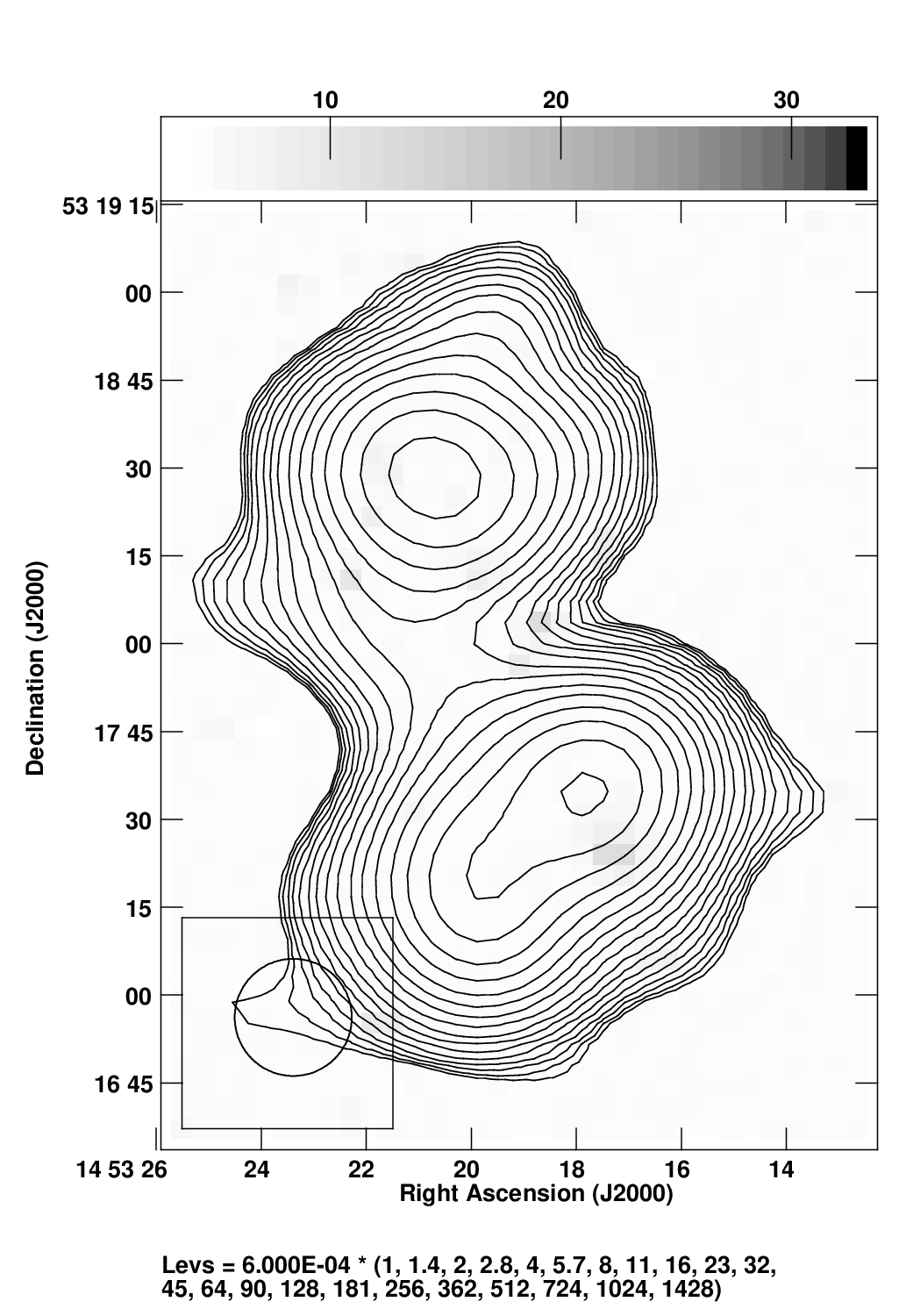}
}
}
\caption{A sample of six LOFAR images of the {\it Z}-shaped radio source (contours) overlaid on the DSS2 red image (grey scale).
The relative radio contours are drawn from a selected lowest level (depending upon the local noise) with a increment of $\sqrt2$.}
\label{fig:sample_zrg}
\end{figure*}

\section{Result}
\label{sec:result}
\subsection{Source Catalogs}
We made a systematic search for the winged radio sources from the LoTSS DR1. A total of 33 winged sources are discovered, out of which 21 are XRGs, and the rest 12 are ZRGs. Example of six XRGs and ZRGs are presented in Figure \ref{fig:sample_xrg} and Figure \ref{fig:sample_zrg}, respectively. In one corner of the images, synthesized beams are shown.

The basic parameters of newly identified XRGs and ZRGs are mentioned in Table \ref{tab:X-shaped} and Table \ref{tab:Z-shaped} respectively. The candidates are catalogued in ascending order of their respective Right Ascension (RA). In columns 3 and 4 of Table \ref{tab:X-shaped} and Table \ref{tab:Z-shaped}, the positions of newly identified XRGs and ZRGs are mentioned where the position of optical host galaxies are used when available. In column 5 of the table, we mentioned the names of the survey from where the optical counterparts are taken. In case of no optical counterparts are available, the position is measured from the eye estimation from the corresponding radio image using the location of the radio core of the intersection of primary and secondary jets. In column 6, red-shift ($z$) is mentioned, when available. In column 7, the flux densities at 144 MHz, using the LoTSS DR1 ($F_{144}$) are mentioned for all the sources and in column 8, corresponding flux densities at 1400 MHz using NRAO/VLA Sky Survey (NVSS, \citet{Co98}) ($F_{1400}$) are mentioned. NVSS is used instead of FIRST because the latter is prone to flux density loss due to the lack of short uv spacing. The flux density measurements are done using Astronomical Image Processing System ({\tt AIPS}) task {\tt TVSTAT}. In column 9, the spectral index ($\alpha_{144}^{1400}$) between 144 and 1400 MHz is tabulated. In columns 10 and 11, luminosity ($L_{rad}$) and power ($P$) of all the sources are tabulated where redshifts are available. In column 12, FR types (whether the primary lobes are FR I or FR II) are mentioned. In column 13, the list of other catalogues is mentioned where these sources are previously mentioned, without identifying them as winged radio galaxies. 
\begin{table*}
\caption{\bf Candidates of Winged Radio Source with {\it X}-shape}
\scriptsize
\tiny
\begin{center}
\begin{tabular}{|c|c|c|c|c|c|c|c|c|c|c|c|c|}
\hline \hline
Cat& Host & R.A.    & Decl.   & Ref. &Redshift&$F_{144}$&$F_{1400}$&$\alpha_{144}^{1400}$&   $L$   &$P$ &FR&Other\\
   &      &         &         &      &        &         &          &                     &  (erg sec$^{-1}$)  & (W Hz$^{-1}$)&type&\\
No.& Name &(J2000.0)&(J2000.0)&      &($z$)  &  (mJy)  & (mJy)    &		        &($\times 10^{42}$)&($\times 10^{25}$)&(I/II) &Catalogs\\
	\hline 
~1	& J1054+5537	& 10 54 51.64	& +55 37 36.3	& SDSS & 0.924	& ~~88	& ~08	& -- 1.05&~5.04	&~41.66& I   &1, 4  \\
~2	& J1056+5111	& 10 56 44.38	& +51 11 59.6	& EE   & ---	& ~~94	& ~19	& -- 0.70& ---	& ---  & I   &1     \\
~3	& J1109+5314	& 11 09 33.92	& +53 14 11.4	& EE   &~0.508$^{\ast}$& ~725	& 114	& -- 0.81& 15.43&~71.61& II  &6     \\
~4	& J1112+4755	& 11 12 17.58	& +47 55 56.6	& SDSS & 0.459	& 2146	& 239	& -- 0.96& 25.37&177.37& --- &1, 2, 3, 6, 7 \\
~5	& J1129+5407	& 11 29 22.09	& +54 07 38.5	& EE   & ---	& ~378	& ~22	& -- 1.25& ---	& ---  & I   &1     \\
~6	& J1132+5558	& 11 32 22.74	& +55 58 18.5	& SDSS & 0.050	& ~695	& ~55	& -- 1.11&~0.05	&~0.45& I   &1, 2  \\
~7	& J1139+5539	& 11 39 08.58	& +55 39 52.1	& SDSS & 0.061	& 1726	& 281	& -- 0.80&~0.38	&~1.69& I   &1, 2, 3, 7\\
~8	& J1154+4835	& 11 54 18.72	& +48 35 21.1	& SDSS &0.170$^{\ast}$& ~665	& 104	& -- 0.82&~1.19	&~5.56& II  &1, 3, 7, 8 \\
~9	& J1201+4925	& 12 01 21.06	& +49 25 14.4	& SDSS &~0.424$^{\ast}$& 2140	& 160	& -- 1.14& 16.98&156.33& --- &1     \\
10	& J1216+5243	& 12 16 23.68	& +52 43 59.9	& SDSS & 0.121	& 210	& ~62	& -- 0.54&~0.53	&~0.83& --- &1, 10 \\
11	& J1228+5527	& 12 28 51.32	& +55 26 54.1	& SDSS &~0.410$^{\ast}$& ~~69	& ~20	& -- 0.54&~2.33	&~3.79& I   &1     \\
12	& J1236+5126	& 12 36 50.27	& +51 26 21.9	& EE   & 0.341	& ~663	& 115	& -- 0.77&~6.44	&~25.88& I   &1, 2  \\
13	& J1148+4653	& 11 48 38.49	& +46 53 11.7	& SDSS & 0.599	& ~~55	& ~08	& -- 0.85& 1.58	&~8.13& I   &	1	\\
14	& J1311+5502	& 13 11 52.51	& +55 02 30.2	& EE   & ---	& ~104	& ~11	& -- 0.99& ---	& ---  & I   &1     \\
15	& J1324+5041	& 13 24 35.20	& +50 41 02.3	& SDSS & 0.287	& ~618	& ~40	& -- 1.20&~1.89	&~18.28& II  &2, 3  \\
16	& J1336+4900	& 13 36 16.03	& +49 00 00.5	& EE   & ---	& ~596	& 112	& -- 0.74& ---	& ---  & II  &1, 7  \\
17	& J1345+5403	& 13 45 57.55	& +54 03 16.6	& SDSS & 0.162	& 3139	& 340	& -- 0.98&~3.42	&~24.52& II  &5, 7  \\
18	& J1414+4842	& 14 14 07.70	& +48 41 56.3	& SDSS & ---	& 1168	& 238	& -- 0.70& ---	& ---  & II  &3     \\
19	& J1416+5425	& 14 16 29.03	& +54 25 32.1	& EE   & ---	& ~983	& ~52	& -- 1.29& ---	& ---  & II  &3     \\
20	& J1518+5237	& 15 18 45.38	& +52 37 08.6	& SDSS & 0.521	& ~303	& ~25	& -- 1.10&~4.08	&~35.82& II  &1     \\
21	& J1442+5043	& 14 42 19.18	& +50 43 57.9	& SDSS & 0.174	& 1140	& 204	& -- 0.76& 2.61	&10.01& I   & 1, 3, 7   \\
\hline
\end{tabular}
\end{center}
References: SDSS: \citep{Fa99, Tu13} and 1: NVSS \citep{Co98}; 2: VLSS \citep{Co07, He08}; 3: 6C \citep{Ba85, Ha88, Ha90, Ha91, Ha93a, Ha93b}; 4: 7C \citep{Mc90, Ko94, Wa96, Ve98}; 5: 8C \citep{Ha95}; 6: TXS \citep{Do96}; 7: 87GB \citep{Gr91}; 8: GB6 \citep{Gr96}; 9: ATATS \citep{Cr10}, 10: VSA \citep{Ta03, Cl08}\\
\scriptsize
\label{tab:X-shaped}
\end{table*}

\begin{figure}
	\centering
\includegraphics[width=7cm,angle=-90,origin=c]{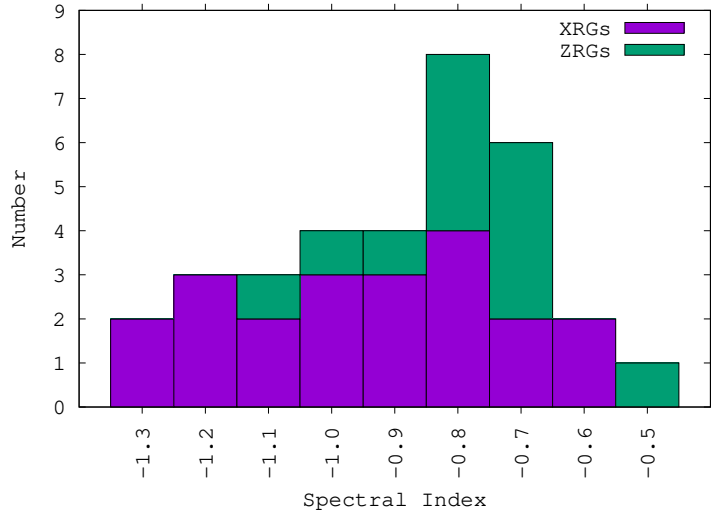}
\caption{Histogram showing spectral index ($\alpha_{144}^{1400}$) distribution of winged radio galaxies presented in current paper.}
\label{fig:spindex_histgrm}
\end{figure}

\subsection{Spectral Index}
The two-point spectral index ($\alpha_{144}^{1400}$) is calculated between 144 and 1400 MHz assuming $S \propto \nu^{\alpha}$, where $\alpha$ is the spectral index and $S_{\nu}$ is the radiative flux density at a given frequency $\nu$. The 144 MHz flux is measured from the LoTSS images and 1400 MHz flux from corresponding NVSS images. These spectral indices ($\alpha_{144}^{1400}$) are determined by measuring flux at both frequencies over the same aperture. These spectral indexes are tabulated in column 9 of Table \ref{tab:X-shaped} and Table \ref{tab:Z-shaped}.  

The uncertainty in calculation of spectral index due to flux density uncertainty \citep{Ma16} is
\begin{equation}
	\Delta\alpha=\frac{1}{\ln\frac{\nu_{1}}{\nu_{2}}}\sqrt{\left(\frac{\Delta S_{1}}{S_{1}}\right)^{2}+\left(\frac{\Delta S_{2}}{S_{2}}\right)^{2}}
\end{equation}
where $S_{1, 2}$ and $\nu_{1, 2}$ are LoTSS DR1 and NVSS flux densities and frequencies respectively. The flux density accuracy in NVSS and LoTSS DR1 is $\sim$5 per cent \citep{Co98} and $\sim$10 per cent \citep{Sh19}. Using equation 1, the spectral index uncertainty is $\Delta\alpha$=0.05.

The radio spectrum of the majority of XRGs and ZRGs is steep ($\alpha_{144}^{1400}>0.5$), which is typical of lobe-dominated radio galaxies. For all of the XRGs presented in the current paper $\alpha_{144}^{1400}>0.5$ and for ZRGs, except one source (J1243+5212) all the sources show a steep spectrum ($\alpha_{144}^{1400}>0.5$). J1243+5212 is the only ZRG that shows a flat spectrum.

A histogram with the spectral index distribution of winged radio galaxies presented in the current article is shown in Figure \ref{fig:spindex_histgrm}. The histogram shows a peak near $\alpha_{144}^{1400}\sim$--0.6 for both XRGs and ZRGs. The total span of spectral index for XRGs is $ 0.54 < |\alpha| < 1.29$and total span of spectral index for ZRGs is $ 0.44 < |\alpha| < 1.02$. The spectral index of XRGs has mean and median values of --0.89 and --0.85, while the spectral index of ZRGs has mean and median values of --0.70 and --0.70, respectively. 

For XRGs, the source with steepest spectrum is J1416+5425 ($|\alpha_{144}^{1400}|=1.29$) while J1216+5243 and J1228+5527 has the flatest spectrum ($|\alpha_{144}^{1400}|=0.54$). For ZRGs, J1126+5334 has the steepest spectrum ($|\alpha_{144}^{1400}|=1.02$) and J1243+5212 has the flatest spectrum ($|\alpha_{144}^{1400}|=0.44$). 

The spectral index of most of the regular radio galaxies lies in the range of --0.7 to --0.8 \citep{Oo88, Ka98, Is10, Ma16} and the average spectral index of XRGs and ZRGs of current paper also lies in the range showing that XRGs and ZRGs are not significantly different than the regular galaxies in terms of the spectral index.

\begin{figure}
  \centering
\includegraphics[width=7.5cm,angle=-90,origin=c]{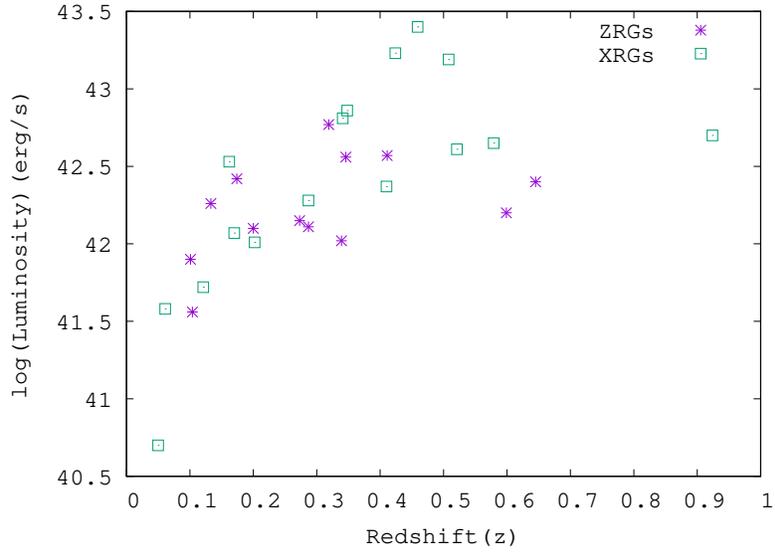}
\caption{Plot showing the distribution of radio luminosity ($L_{rad}$) with redshift ($z$) for both the XRGs and ZRGs.}
\label{fig:luminosity_redshift}
\end{figure}

\subsection{Radio Luminosity and Power}
The Radio luminosity {$L_{rad}$} of XRGs and ZRGs presented in the current paper is calculated using
\begin{equation}
\begin{split}
L_{rad} &= 1.2\times10^{27}D^2_{\text{Mpc}}S_0{\nu_0^{-\alpha}}(1+z)^{-(1+\alpha)}          \\
        &  \times(\nu_u^{(1+\alpha)}-\nu_l^{(1+\alpha)})(1+\alpha)^{-1} \text{erg sec$^{-1}$} \\
\label{eqn:1}
\end{split}
\end{equation}
where $D_{\text{Mpc}}$ is the luminosity distance to the source in Mpc, $z$ is the redshift of the radio galaxy, $S_0$ is the flux density (in Jy) at a given frequency $\nu_0$ (in Hz), $\alpha$ is the spectral index ($S \propto \nu^\alpha$), and $\nu_l$ (Hz) and $\nu_u$ (Hz) are the lower and upper cut-off frequencies \citep{Od87}.
We consider the lower and upper cut-off frequencies as 10 MHz and 100 GHz respectively.

The radio luminosity is evaluated for all 26 sources (14 XRGs and 12 ZRGs) for which redshift was available. At 144 MHz, the source radio luminosities are in the order of $10^{42}$ erg sec$^{-1}$, which is similar to a normal radio galaxy. Amongst the XRGs presented in the current paper, J1112+4755 has the maximum luminosity with $\log L=43.40$ erg sec$^{-1}$ and J1132+5558 has the minimum luminosity with $\log L=40.70$ erg sec$^{-1}$. For ZRGs, J1453+5318 has the maximum luminosity with $\log L=42.77$ erg sec$^{-1}$ and J1121+5344 has the minimum luminosity with $\log L=41.56$ erg sec$^{-1}$.

In Figure \ref{fig:luminosity_redshift}, distribution of $\log$ of radio luminosity ($\log L_{rad}$) with redshift ($z$) is presented. It is found that $\sim$70$\%$ of the sources have the luminosity in the order of $10^{42}$ erg sec$^{-1}$. For the XRGs the mean and median of $\log L$ are 42.39 erg sec$^{-1}$ and 42.42 erg sec$^{-1}$ respectively. The mean and median of $\log L$ for ZRGs are 42.23 erg sec$^{-1}$ and 42.15 erg sec$^{-1}$ respectively. The mean $\log L$ of XRGs is slightly high compared to ZRGs, as was found previously for the cases of winged radio galaxies from the FIRST \citep{Be20} and TGSS \citep{Bh20}.

\begin{figure}
  \centering
\includegraphics[width=7cm,angle=-90,origin=c]{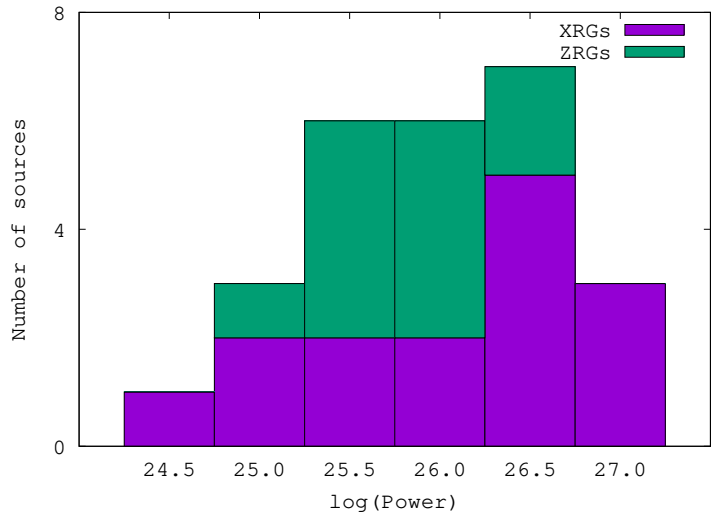}
\caption{Histogram showing the distribution of $\log P$ for winged sources presented in the current paper.}
\label{fig:power_histogram}
\end{figure}

The radio power is calculated using the following formula, given by \cite{Do09}
\begin{equation}
P_{\nu} = 4{\pi}D^2_{L}S_{\nu}(1 + z)^{({\alpha} - 1)}
\\\label{eqn:2}
\end{equation}
where $D_{L}$ is the luminosity distance, $S_{\nu}$ is the radio flux at a frequency $\nu$, $(1 + z)^{({\alpha} - 1)}$ is the standard k-correction used in radio astronomy.
In this equation, the spectral index ($\alpha$) is assumed to follow as $S_{\nu} \propto \nu^{-\alpha}$.
The calculated powers for all XRGs and ZRGs are presented in column 11 of Table \ref{tab:X-shaped} and Table \ref{tab:Z-shaped}.
For all winged radio galaxies, $\log P$ lies in the range 24.65 to 27.25.
Amongst XRGs, J1112+4755 has the highest power with $\log P$=27.25 and J1132+5558 has the lowest power with $\log P$=24.65.
Similarly, amongst ZRGs, J1453+5318 has the maximum power with $\log P$=26.35 and J1243+5212 has the minimum power with $\log P$=25.10.
The mean and median of $\log P$ for XRGs are 26.10 and 26.26 respectively and for ZRGs the mean and median are 25.78 and 25.80 respectively. The overall mean and median $\log P$ of the winged sources in the current paper are 25.97 and 25.95 respectively.
XRG sources have slightly more average power compared to ZRGs. A histogram of $\log P$ is presented in Figure \ref{fig:power_histogram}. There is a peak close to $\log P\sim$26.5 for winged radio galaxies presented in the current paper.

\begin{figure}
  \centering
\includegraphics[width=7.5cm,angle=-90,origin=c]{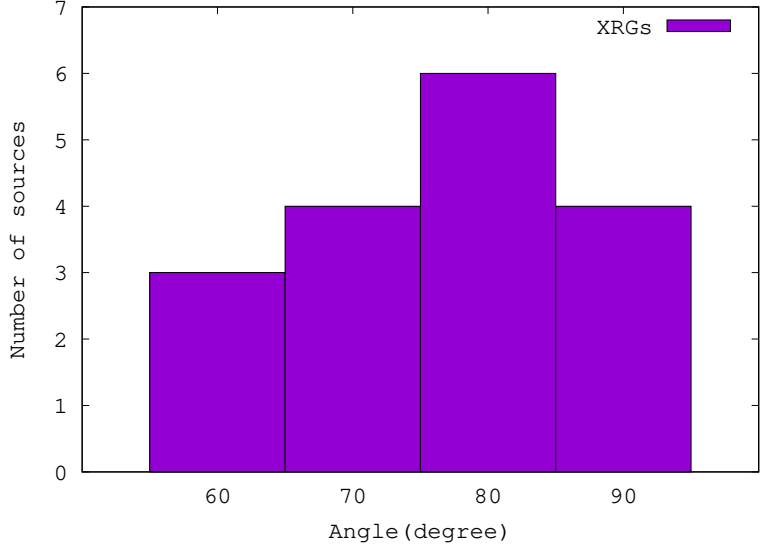}
\caption{Histogram showing the distribution of angles between the major and minor axis for XRGs.}
\label{fig:angle_histogram}
\end{figure}

\begin{table*}
\caption{\bf Candidates of Winged Radio Source with {\it Z}-shape}
\scriptsize
\tiny
\begin{center}
\begin{tabular}{|c|c|c|c|c|c|c|c|c|c|c|c|c|}
\hline \hline
Cat& Host & R.A.    & Decl.   & Ref. &Redshift&$F_{144}$&$F_{1400}$&$\alpha_{144}^{1400}$&   $L$   &$P$ &FR&Other\\
   &      &         &         &      &        &         &          &                     &  (erg sec$^{-1}$)  & (W Hz$^{-1}$)&type&\\
No.& Name &(J2000.0)&(J2000.0)&      &($z$)  &  (mJy)  & (mJy)    &		        &($\times 10^{42}$)&($\times 10^{25}$)&(I/II) &Catalogs\\
\hline
~1	& J1121+5344	& 11 21 26.44	& +53 44 56.7	& SDSS & 0.104	& ~697  & ~95	& -- 0.88& 0.36	&~2.04& II  & 3	\\
~2	& J1126+5334	& 11 26 39.89	& +53 34 27.1	& SDSS & 0.645	& ~102	& ~10	& -- 1.02& 2.50	&19.54& II  & 1	\\
~3	& J1228+5431	& 12 28 38.39	& +54 31 39.5	& SDSS & 0.339	& ~169	& ~20	& -- 0.94& 1.04	&~6.81& I   & 1	\\
~4	& J1243+5212	& 12 43 08.97	& +52 12 45.0	& SDSS & 0.200	& ~112	& ~41	& -- 0.44& 1.25	&~1.26& I   & 1, 8	\\
~5	& J1314+5439	& 13 14 04.60	& +54 39 37.9	& SDSS & 0.346	& ~394	& ~65	& -- 0.79& 3.67 &15.92& --- & 1, 7	\\
~6	& J1315+5254	& 13 15 28.82	& +52 54 40.0	& SDSS &~0.411$^{\ast}$& ~203	& ~41	& -- 0.70& 3.76	&11.85& I   & 1, 7, 9   \\
~7	& J1346+5410	& 13 46 32.69	& +54 10 31.6	& SDSS &~0.273$^{\ast}$& ~265	& ~43	& -- 0.80& 1.42	&~6.32& II  & 1, 9	\\
~8	& J1446+4832	& 14 46 26.09	& +48 32 04.4	& SDSS & ---	& ~177	& ~31	& -- 0.76& ---	& --- & --- & 9	\\
~9	& J1453+5318	& 14 53 19.08	& +53 17 55.8	& SDSS &~0.319$^{\ast}$& ~676	& 121	& -- 0.76& 5.88	&22.54& II  & 2, 3, 6, 7   \\
10	& J1502+5304	& 15 02 09.53	& +53 04 19.9	& SDSS & 0.287	& ~147	& ~31	& -- 0.68& 1.30	&~3.81& --- & 1, 9	\\
11	& J1502+5244	& 15 02 29.04	& +52 44 02.1	& SDSS & 0.133	& ~820	& 205	& -- 0.61& 1.83	&~3.94& II  & 1, 2, 7, 9  \\
12	& J1239+5314	& 12 39 15.39	& +53 14 14.6	& SDSS & 0.202	& 264	& ~54	& -- 0.70&~1.02	&~3.15& II  &7     \\								
\hline
\end{tabular}
\end{center}

References: SDSS: \citep{Fa99, Tu13} and 1: NVSS \citep{Co98}; 2: VLSS \citep{Co07, He08}; 3: 6C \citep{Ba85, Ha88, Ha90, Ha91, Ha93a, Ha93b}; 4: 7C \citep{Mc90, Ko94, Wa96, Ve98}; 5: 8C \citep{Ha95}; 6: TXS \citep{Do96}; 7: 87GB \citep{Gr91}; 8: GB6 \citep{Gr96}; 9: ATATS \citep{Cr10}, 10: VSA \citep{Ta03, Cl08}\\
\scriptsize
\label{tab:Z-shaped}
\end{table*}

\section{Some morphological properties of X-shaped radio galaxies}
\subsection{Symmetry}
For different XRGs, the wings originated from the primary jets are different in shape, size, point of ejection, and orientation of jets. For all of the 21 XRGs reported in the current paper, both of the wings are different in either shape or size or alignment or more than one such condition and these sources are asymmetric in nature. Out of the 21 XRGs, 4 sources have a one-sided wing. For ZRGs in the current paper, all the sources have an asymmetric structure.

\subsection{Angle between primary and secondary lobes}
We estimated the angle between primary and secondary lobes of XRGs by drawing two straight lines along the major and minor axis. The range of the angles is 54 degrees to 88 degrees. The lowest angle of 54 degrees is found for the source J1109+5314. The source J1324+5041 has an angle of 88 degrees between the major and minor axis.

A histogram for the angles is presented in Figure \ref{fig:angle_histogram}. There is a peak near 80--90 degrees in the histogram.

\section{Discussion}
\label{sec:disc}
In this paper, we present the systematic search result for the winged radio galaxies from LoTSS DR1. We detected 33 winged sources out of which 21 are XRGs and 12 are ZRGs.

We found redshift for 26 ($\sim$79 per cent of total) winged radio galaxies presented in the current paper. All the winged sources have the redshift value in the range of 0.050 to 0.924. For XRGs, redshift has been found for $\sim$ 71 per cent of the sources. Among XRGs in the current sample, the source J1054+5537 is the furthest one with a redshift 0.924 \citep{Al15} and J1132+5558 is the nearest one with redshift 0.050 \citep{Al15}. We found redshift for all of the ZRGs, except the source J1446+4832. Out of the twelve ZRGs, J1126+5334 is the furthest one with redshift 0.645 \citep{Al15} and J1121+5344 is the closest one with redshift 0.104 \citep{Al15}.

All of our winged radio sources (except one) show a steep radio spectrum and this is expected for lobe-dominated radio sources. Among XRGs 8 sources have FR-II type, 10 sources have FR-I type morphology and 3 sources have mixed morphology. Among ZRGs 6 sources have FR-II type, 3 sources have FR-I type morphology and 3 sources have mixed morphology. So, we found more FR-II (14) than FR-I (13) sources. The mean and median $\log L$ for XRGs in the present paper is 42.39 erg sec$^{-1}$ and 42.42 erg sec$^{-1}$ respectively. For ZRGs, the mean and median $\log L$ are 42.23 erg sec$^{-1}$ and 42.15 erg sec$^{-1}$ respectively. The mean and median $\log L$ of winged sources presented in the current paper are a little low compared to the average $\log L$ (=44.07 erg sec$^{-1}$) calculated by \citet{Ni93}. We computed the mean $\log P$ of sources presented in the current paper as 26.03 W Hz$^{-1}$ which is close compared with the power evaluated by \cite{Da20} for the LOFAR sources. Among all winged radio galaxies, J1112+4755 is the brightest XRGs with a flux density $F_{144}=2146$ mJy and J1502+5244 is the brightest ZRGs with a flux density $F_{144}=820$ mJy.

The reason for wings in XRGs and ZRGs are still not known, though there are a lot of models. More identification of such sources and multi-wavelength study of this kind of source along with their characteristic study will lead to a better understanding of this interesting class of radio sources.

\section{Conclusion}
\label{sec:conc}
A list of 33 winged radio galaxies from LoTSS DR1 is prepared, among them 21 are XRGs, and 12 are ZRGs. Finding these sources increases the number of known XRGs and ZRGs. We calculated various physical properties of these sources like spectral index, luminosity, and power. The optical counterparts are found for twenty-six sources. The mean and median $\log L$ value of XRGs is less than that of ZRGs. A deep multi-wavelength follow-up observation is required for understanding the exact nature of these sources.

\section*{Acknowledgments}
The authors would like to acknowledge the NASA/IPAC Extra-galactic Database (NED) which is operated by the Jet Propulsion Laboratory, California Institute of Technology, under contract with the National Aeronautics and Space Administration.



\begin{thebibliography}{99}
\bibitem[{Aghanim et al.}(2020)]{Ag20}
Aghanim N. et al., 2020, A\&A, 641, 67

\bibitem[{Alam et al.}(2015)]{Al15}
Alam S. et al., ApJS, 2015, 219, 12A

\bibitem[{Baldwin et al.}(1985)]{Ba85}
Baldwin J. E., Boysen R. C., Hales S. E. G., Jennings J. E., Waggett P. C., Warner P. J., Wilson D. M. A., 1985, MNRAS, 217, 717

\bibitem[{Bera et al.}(2020)]{Be20}
Bera S., Pal S., Sasmal T. K., Mondal S., 2020, ApJS, 251, 15

\bibitem[{Bhukta, Pal \& Mondal}(2020)]{Bh20}
Bhukta N., Pal S., Mondal S., 2020, MNRAS, submitted, arXiv:2006.07219

\bibitem[{Capetti et al.}(2001)]{Ca02}
Capetti A., Zamfir S., Rossi P., Bodo G., Zanni C., Massaglia S., 2002, A\&A, 394, 39

\bibitem[{Cleary et al.}(2008)]{Cl08}
Cleary K. A. et al., 2008, MNRAS, 386, 1759

\bibitem[{Cohen et al.}(2007)]{Co07}
Cohen A. S., Lane W. M., Cotton W. D., Kassim N. E., Lazio T. J. W., Perley R. A., Condon J. J., Erickson W. C., 2007, AJ, 134, 1245

\bibitem[{Condon et al.}(1998)]{Co98}
Condon J. J., Cotton W. D., Greisen E. W., Yin Q. F., Perley R. A., Taylor G. B., Broderick J. J., 1998, AJ, 115, 1693

\bibitem[{Cheung}(2007)]{Ch07}
Cheung C. C., 2007, ApJ, 133, 2097

\bibitem[{Croft et al.}(2010)]{Cr10}
Croft S. et al, 2010, AJ, 719, 45

\bibitem[{Dabhade et al.}(2020)]{Da20}
Dabhade P. et al., 2020, A\&A, 635, A5

\bibitem[{Dennett-Thorpe et al.}(2002)]{De02}
Dennett-Thorpe J., Scheuer P. A. G., Laing R. A., Bridle A. H., Pooley G. G., Reich W., 2002, MNRAS, 330, 609

\bibitem[{Donoso, Best \& Kauffmann}(2009)]{Do09}
Donoso E., Best P. N., Kauffmann G., 2009, MNRAS, 392, 617

\bibitem[{Douglas et al.}(1996)]{Do96}
Douglas J. N., Bash F. N., Bozyan F. A., Torrence G. W., Wolfe C., 1996, AJ, 111, 1945

\bibitem[{Ekers et al.}(1978)]{Ek78}
Ekers R. D., Fanti R., Lari C., Parma, P., 1978, Nature, 276, 588

\bibitem[{Fan et al.}(1999)]{Fa99}
Fan X. et al., 1999, AJ 118, 1

\bibitem[{Fanaroff \& Riley}(1974)]{Fa74}
Fanaroff B. L., Riley J. M., 1974, MNRAS, 167, 31P

\bibitem[{Ficarra, Grueff \& Tomassetti}(1985)]{Fi85}
Ficarra A., Grueff G., Tomassetti G., 1985, A\&A, 59, 255

\bibitem[{Gopal-Krishna, Biermann \& Witta}(2003)]{Go03}
Gopal-Krishna, Biermann P. L., Wiita P. J., 2003, ApJ, 594, L103

\bibitem[{Gregory \& Condon}(1991)]{Gr91}
Gregory P. C., Condon J. J., 1991, ApJS, 75, 1011

\bibitem[{Gregory et al.}(1996)]{Gr96}
Gregory P. C., Scott W. K., Douglas K., Condon J. J., 1996, ApJS, 103, 427

\bibitem[{Gull \& Northover}(1973)]{Gu73}
Gull S. F., Northover K. J. E., 1973, Nature, 244, 80

\bibitem[{Gunn et al.}(2006)]{Gu06}
Gunn J. E. et al., 2006, AJ, 131, 2332

\bibitem[{Hales, Baldwin \& Warner}(1988)]{Ha88}
Hales S. E. G., Baldwin J. E., Warner P. J., 1988, MNRAS, 234, 919

\bibitem[{Hales et al.}(1990)]{Ha90}
Hales S. E. G., Masson C. R., Warner P. J., Baldwin J. E., 1990, MNRAS, 246, 256

\bibitem[{Hales et al.}(1993a)]{Ha93a}
Hales S. E. G., Baldwin J. E., Warner P. J., 1993a, MNRAS, 263, 25

\bibitem[{Hales et al.}(1993b)]{Ha93b}
Hales S. E. G., Masson C. R., Warner P. J., Baldwin J. E., Green D. A., 1993b, MNRAS, 262, 1057

\bibitem[{Hales et al.}(1991)]{Ha91}
Hales S. E. G., Mayer C. J., Warner P. J., Baldwin J. E., 1991, MNRAS, 251, 46

\bibitem[{Hales et al.}(1995)]{Ha95}
Hales S. E. G., Waldram E. M., Rees N., Warner P. J., 1995, MNRAS, 274, 447

\bibitem[{Heald et al.}(2015)]{He15}
Heald G. H. et al. 2015, A\&A, 582, A123

\bibitem[{Helmboldt et al.}(2008)]{He08}
Helmboldt J. F., Kassim N. E., Cohen A. S., Lane W. M., Lazio T. J., 2008, ApJS, 174, 313

\bibitem[{Hodges-Kluck \& Reynolds}(2011)]{Ho11}
Hodges-Kluck E. J, Reynolds C. S, 2011, ApJ, 733, 58

\bibitem[{Intema et al.}(2017)]{In17}
Intema H. T., Jagannathan P., Mooley K. P., Frail D. A., 2017, A\&A, 598, A78

\bibitem[{Ishwara-Chandra et al.}(2010)]{Is10}
Ishwara-Chandra C. H. et al., 2010, \mnras, 405, 436

\bibitem[{Kapahi et al.}(1998)]{Ka98}
Kapahi V. K. et al., 1998, \apjs, 118, 275

\bibitem[{Kollgaard et al.}(1994)]{Ko94}
Kollgaard R. I., Brinkmann W., Chester M. M., Feigelson E. D., Hertz P., Reich P., Wielebinski R., 1994, ApJS, 93, 145

\bibitem[{Lal \& Rao}(2005)]{La05}
Lal D. V., Rao A. P., 2005, MNRAS, 356, 232

\bibitem[{Lal \& Rao}(2007)]{La07}
Lal D. V., Rao A. P., 2007, MNRAS, 374,1085

\bibitem[{Lane et al.}(2014)]{La14}
Lane W. M. et al., 2014, MNRAS, 440, 327

\bibitem[{Leahy \& Parma}(1992)]{Le92}
Leahy J. P., Parma P., 1992, in Extragalactic Radio Sources: From Beams to Jets, ed. J. Roland H. Sol, G. Pelletier (Cambridge: Cambridge Univ.Press), 307

\bibitem[{Leahy \& Williams}(1984)]{Le84}
Leahy J. P., Williams A. G., 1984, MNRAS, 210, 929

\bibitem[{Mahony et al.}(2016)]{Ma16}
Mahony E.K. et al., 2016, \mnras, 463, 2997

\bibitem[{McGilchrist et al.}(1990)]{Mc90}
McGilchrist M. M., Baldwin J. E., Riley J. M., Titterington D. J., Waldram E. M., Warner P. J., 1990, MNRAS, 246, 110

\bibitem[{Merritt \& Ekers}(2002)]{Me02}
Merritt D., Ekers R. D., 2002, Science, 297, 1310

\bibitem[{Mingo et al.}(2019)]{Mi19}
Mingo et al., 2019, MNRAS, 488, 2701

\bibitem[{Nilsson et al.}(1993)]{Ni93}
Nilsson K., Valtonen M. J., Kotilainen J., Jaakkola T., 1993, ApJ, 413, 453

\bibitem[{O'Dea \& Owen}(1987)]{Od87}
O'Dea C. P., Owen F. N., 1987, AJ, 316, 95

\bibitem[{Oort, Steemers \& Windhorst}(1988)]{Oo88}
Oort M. J. A., Steemers W. J. G., Windhorst R. A., 1988, A\&AS, 73, 103

\bibitem[{Pal \& Kumari}(2021)]{Pa21}
Pal S., Kumari S., 2021, submitted, arXiv:2103.15153

\bibitem[{Proctor}(2011)]{Pr11}
Proctor D. D., 2011, ApJS, 194, 33

\bibitem[{Riley}(1972)]{Ri72}
Riley J. M., 1972, MNRAS, 157, 349

\bibitem[{Rottmann}(2001)]{Ro01}
Rottmann, H. 2001, PhD thesis , Univ. Bonn

\bibitem[{Shimwell et al.}(2017)]{Sh17}
Shimwell T. W. et al., 2017, A\&A 598, A104

\bibitem[{Shimwell et al.}(2019)]{Sh19}
Shimwell T. W. et al., 2019, A\&A 622, A1

\bibitem[{Taylor et al.}(2003)]{Ta03}
Taylor A. C. et al., 2003, MNRAS, 341, 1066

\bibitem[{Tumlinson et al.}(2013)]{Tu13}
Tumlinson J. et al., 2013, ApJ, 777, 33

\bibitem[{Vessey \& Green}(1998)]{Ve98}
Vessey S. J., Green D. A., 1998, MNRAS, 294, 607

\bibitem[{Waldram et al.}(2003)]{Wa03}
Waldram E. M., Pooley G. G., Grainge K. J. B., Jones M. E., Saunders R. D. E., Scott P. F., Taylor A. C., 2003, MNRAS, 342, 915

\bibitem[{Waldram et al.}(2010)]{Wa10}
Waldram E. M., Pooley G. G., Davies M. L., Grainge K. J. B., Scott P. F., 2010, MNRAS, 404, 1005

\bibitem[{Waldram et al.}(1996)]{Wa96}
Waldram E. M., Yates J. A., Riley J. M., Warner P. J., 1996, MNRAS, 282, 779

\bibitem[{Wayth et al.}(2015)]{Wa15}
Wayth R. B. et al., 2015, PASA, 32, e025

\bibitem[{Williams}(1991)]{Wi91}
Williams A., 1991, in Beams and Jets in Astrophysics, ed. Huges P., Cambridge University Press, 1991, 342 

\bibitem[{Yang et al.}(2019)]{Ya19}
Yang et al., 2019, ApJS, 245, 17

\bibitem[{Zier}(2005)]{Zi05}
Zier C., 2005, MNRAS, 364, 583

\end{thebibliography}
\end{document}